# On the Software and Knowledge Engineering Aspects of the Educational Process

THANASSIS HADZILACOS[1,2], DIMITRIS KALLES[1], and MARIA POULIOPOULOU[1]

[1] *Hellenic Open University,*
*Laboratory of Educational Material and Educational Methodology,*
*Sachtouri 23, 26222, Patras, Greece*

[2] *Research Academic Computer Technology Institute, Patras, Greece*
*Contact: kalles@eap.gr*

ABSTRACT

The Hellenic Open University has embarked on a large-scale effort to enhance its textbook-based material with content that demonstrably supports the basic tenets of distance learning. The challenge is to set up a framework that allows for the production-level creation, distribution and consumption of content, and at the same time, evaluate the effort in terms of technological, educational and organizational knowledge gained. This paper presents a model of the educational process that is used as a development backbone and argues about its conceptual and technical practicality at large.

## 1 The educational process as a complex system

Depending on how one views the educational process there are distinct components of it which become eminent during the observation. Even if each observer does in fact glimpse all components of the process, the emphasis is always on some key ones.

A teacher, for example, usually views the educational process as a set of lectures to be delivered to an audience. Peripheral aspects of this view concern the distribution and grading of assignments and examinations.

A learner, on the other hand, may or may not attend lectures. Attending lectures is only one of the activities that the learner has at his disposal. Studying, experimenting and collaborating are all activities that help hone a skill or develop knowledge about a subject. Informal communication and collaboration among peers is a key aspect of a learner's activities that a teacher may have little, if any, influence. In such a collaboration context, views and homework solutions can be exchanged. Unless the teacher has explicitly designed an assignment to stimulate such communication, the indirect learning effects of the peer collaboration arrive by luck rather than by design.

If we consider teachers and learners to operate at roughly the same level of education, we can move up one level and consider the educational system view. At that view, one deals with providing the educational material at a suitable scale for the student population and setting and monitoring quality issues in the delivery of education (i.e. scope of educational activities, depth and breadth of material, academic prerequisites across subjects, attendance logistics, etc.). Note that, at that level, the delivery mode of education (on-line, physical presence, etc.) is simply another component of that view.

Going a level down from teachers and learners one deals with educational material per se (books, instruments, software, etc.) and the development of blueprints or guides for using that material (solution manuals, demo software activities, etc.). At that level one would also address infrastructure issues.

For each of the above four views (and it should be obvious that the list is not exhausted here), it would be difficult to argue that they are unrelated. These views are not (and should not be) orthogonal, but they help focus the attention of people active in each level towards a common background of experience, expectations, and norms that allows for the smooth exchange of information within the boundaries of that view and across views. This trait of views directly puts them at the centre of both software and knowledge engineering disciplines.

The rest of this paper is structured in five sections. We first develop the rationale for using views to describe complex systems and then argue why we believe a Learner's ODL (Open-and-Distance-Learning) course is a useful generalization of important aspects of the educational process. We then reflect on tools' families that would be fit for the various stakeholders of the educational process. Then, we move to offer some guidelines and issues that need to be answered in an implementation context and conclude by presenting the targeted application domain, which is the mass-scale development of on-line material for a large university.

## 2  Views for complex systems

Identifying views within a process is a key to reducing its complexity and the human race has a long history of doing so.

When designing an artifact ancient craftsmen relied on their mastery and experience to deliver the final product. When artifacts became too large or production shifted from individuals to groups (and, then, to factories), division of labor usually meant that specialization occurred. Within the limits of specialization, practices and terminology emerged that would allow people who practiced an art to convey messages efficiently.

At some point artifacts became too large to easily communicate their complete models in verbatim or prose. Still, complexity had to be kept under control. Some craftsmen began to abstract away from the full description. For example, ship builders stopped using miniature ships as a guide for the workers at the shipyard. Instead, they started using diagrams that would communicate the dimensions and the compartments of the ship. Of course, these diagrams required that shipyard staff would be able to read them: more importantly, they required that consensus existed for what was *not* included in the diagram. Thus formed a subset of people in that industry who were able to communicate effectively and efficiently among them, but still were able to work within the whole production chain, which spanned from unskilled workers to ship-owners.

Within this short example lies perhaps the fundamental tenet of designing by views: views should be simpler than the whole system and consistent with it.

Today, civil engineering can be considered among the most mature of the building industries, at least as far as views are concerned. We all know how to spot a door in an apartment drawing: regardless of scale, language, measurement systems or else, we have come to be able to understand a drawing as a model view of a building and we know that we must look for rooms, windows, doors, etc. and are able to tell these components apart. Other disciplines are following suit, slowly but steadily capitalizing on the fact that successful views manage to limit complexity without sacrificing the final result.

Every view of a complex system is targeted at some stakeholder. A stakeholder is a person, group of people, organization, or any other (suitably defined) entity, for which a particular view is convenient, useful or essential. Also, in principle, each view is an abstraction of a common-for-all underlying model (the final ship, the building, the operating system). This model may logically exist per se, or it may be inferred by integrating the individual views.

It is important to acknowledge the system stakeholders but supporting all of them in a design problem is a different task altogether, especially since it may be too difficult to agree on the underlying model. Some groups are served by highly-focused tools (for example, CAD users or logistics personnel). Maintaining consistency among working groups when the final system is inferred as a sum-of-parts (views), as opposed to a complete design serving as the reference base, may be very difficult. This will be usually due to the tendency to replicate data across applications.

As an example, consider a timber processing industry. Data about the producer of timber X may be useful to engineers who want to keep track on quality issues and specifications. The same data may be useful to logistics because of transport costs. Identifying the need for common data is usually a huge problem in itself and more often than not observed *a posteriori*, after custom systems with inconsistent data formats and interpretations have been developed and fielded.

Data is always tied to some application. Therefore, data replication across applications proves that there exists some common ground between business areas, which may seem unrelated at first sight. If we can confidently pin-point data interchange areas and analyze data usage trends, we can then extract knowledge about the modeled process. This has been practiced and widely publicized from a very narrow point of view to-date: web mining in web site generic usage analysis.

The approaches to integrating views *a posteriori* range in practice, but the underlying concepts have given rise and prominence to the relatively recent field of ontologies [1]. As in earlier approaches [2; and references therein[1]], it seems that the best way to tackle the problem is to analyze independent views and come up with a generic enough model that encompasses them, allowing for object translations between views [3].

---

[1] It is surprising to see that Google Scholar reports that the latest citation to [2] occurs in 1999, a year where the ontology field seems to have started commanding enough attention on its own.

In this paper, we contend that a conceptual artifact, termed a **Learner's ODL course**, is a generic enough model that is suitable for accommodating the most important working practices of the educational model, while still allowing room for developing new abstractions and still be supported by rudimentary technology. We are careful to note that the educational process comprises of observable and explicitly initiated activities, as opposed to the learning process which is ad hoc and may or may not be a direct or indirect outcome of the educational process. After all, education does not necessarily result in measurable learning.

## 3   Using a Learner's ODL course to extend the book

Developing an educational experience for a learner has at least two cornerstones: the existence of educational material and the organization of activities with that material. Meaningful educational experiences are usually based on the shrewd organization of carefully designed activities on quality educational material. The shrewd organization and the careful design necessarily cover some aspects of resource planning, such as how much time the learner is supposed to dedicate to the activity or, what is the sequence of activities that will best attain the educational goal. They also cover some conventional aspects of design, such as the target audience (sophisticated learners or beginners) or, the combination of tools (software, equipment, simulations, etc.) to attain the goal. That way, educational experiences can be then turned into educational material themselves.

A model imposes common language, organization, and disciplines in a community. It is also an excellent guide for tool development, since one will usually want to develop a tool that caters to the needs of a particular group of stakeholders in that community. One may develop functionalities that a group of users deems essential based on these common principles; it will then follow directly where bridging points must exist for the accommodation of other groups of users. These bridging points will necessarily correspond to model components that are shared between views.

Traditional educational activities employ paper based books. This results in several obvious but very important observations which follow from the static nature of textbooks:

- Students treat their books as read-only (apparently so). Moreover, unless authors are talented, books also determinedly convey the read-only message ("you shall not challenge the textbook").
- Authors treat their books as write-once (at best, they edit them if the books are successful and put out a new edition after several years).
- Books are printed and bound sequentially, so they are a sequential medium. Explicit provision is necessary for alternative paths within a textbook (usually, a couple of paragraphs in the preface). The most successful alternative paths are demonstrated in reference books (dictionaries, thesauri, encyclopedias).
    - o  Alternative paths are usually meant as "targeted to graduates" or "targeted to mathematically-inclined readers". In this context,

alternative paths scarcely exist for the much more easily identified student types of "I do not understand this paragraph", "I liked this particular section; which reference should I review?" (and, when they do, they are usually not that specific but still very valuable).
- From an informatics point of view, a textbook is both "data" and "presentation guide". As a field, we have come to a widely held belief that these two should be orthogonal.
- Last, but not least, a textbook reflects the author's views and not the actions that the learner has to perform while learning.

For example, in a conventional textbook, there is a mode for reading the text and there is a mode for selectively browsing the content in search for what will aid us in solving an exercise. Both activities make use of the same underlying asset, in a different fashion.

A simple generalization of the textbook is to make "data" and "presentation guide" orthogonal. Digital platforms allow us to decouple presentation from data. That way, the sequential presentation of the educational material is simply one of the (numerous) available compilations of that material.

A further generalization refers to allowing the book to be re-written, by the author and by the students, while it is being studied and used. This suggests that efficient material updating may have a much more profound effect in the learning practice. Admittedly, today's electronic environments at least offer a variety of updating, indexing and searching mechanisms.

A still further generalization refers to allowing the learning community to come together and to allow tutors to respond dynamically to learners' needs, possibly by discussing and possibly by providing feedback and additional material that will be readily available for the next learners who take that course. This is not exotic technology; most conventional web sites of university courses already sport this feature, whereas learning communities resort to asynchronous discussion forums for interaction ("what did my co-learners think of that exercise", "how do I inform my students about a new tool that supports their homework", "here's an exercise: spend as long as you wish now, but make sure you can do it in five minutes before taking the exam" could be typical comments attached to material and learning activities).

A subtler generalization refers to allowing the book to be used as a notepad, where scratch notes are taken, exercises are attempted, solutions are formed and tests are taken. But, note that, also from a social point of view, we consider a printed book as something that should be preserved as is; in students' used book exchanges, for example, higher prices are commanded for unused books.[2]

However, all the above generalizations can be in principle supported by digital technologies. If, using today's pervasive metaphor, we can envision a book as a web site, we can support all the above features, if we take up the challenge to deal with the complexity of building such a rich-content, rich-interaction web site.

---

[2] This happens for good reason! A densely commented book is of much higher educational value, yet the comments will usually be only appreciated by the student who took them down at the first place.

Extending the book, as set out above, is the core concept of the Learner's ODL course. This paper shows that this is built around mature practices, concepts and technologies. However, it is left to us to glue all these components in a framework that makes sense for all (or, at least, the most important) aspects of the educational process. To see why it is intuitively appealing to merge these concepts, we now delve into a "visual" user-based exploration of the generalizations set out here.

## 4  A Learner's ODL course requirements for view integration

Let us start with a tutor preparing an educational activity. A tutor needs a workbench, where available materials and tools can be combined to create a module that addresses some specific course requirements (for example, instruction, issuing of exercises and homework, etc.).

Figure 1 shows a relatively comprehensive, yet intuitive interface for the tutor's workbench. He will be able to draw together texts and exercises (and other assets) from a suitably organized collection, and he will be able to bundle them together in the context of a learning activity, allowing for comments on how long a student is expected to spend on a particular item. This is a most important stage in distance learning from a tutor's point of view. Notwithstanding the ability of a talented teacher to address the pace of a physical classroom, it is also a hallmark of good preparation in any tutoring context. Up to this point it is rather unimportant whether this workbench is physical (pen and paper and books) or digital, save for the indexing and searching facilities that a digital artefact offers.

At this point a small detour is worthwhile. We stress the importance of designing learning activities that place emphasis on what the learner is expected to do. We explicitly stress that, although this workbench can be used to simply present compilations of material, it is the structuring of the intended usage of the material that creates the learning added value. In this sense, we concur with the view that technology is an excellent supporting tool for education, only when didactic issues are carefully addressed [4, 5]. However, we also stress that our model is capable of supporting both tutor-oriented and learner-oriented work. We claim that this is a result of the decision to first model the process as opposed to first designing the interaction; by setting the model we may then explore interaction designs that are consistent with a master plan.

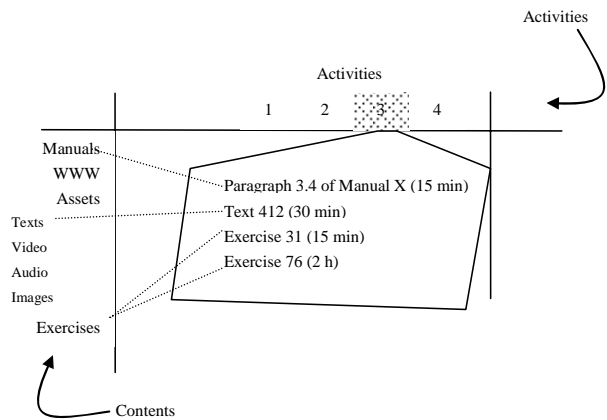

Figure 1: A tutor's basic workbench.

It is interesting to note that the above workbench also relates directly to a curriculum designer, who may prepare a collection of activities based on the collection of assets and on the feedback received by collaborating tutors and students. This feedback is three-fold: based on critique and comments about specific material and assets, based on tutor-student interaction (also allowing tutor-tutor and student-student communication) and based on new material that a tutor adds to the collection of assets.

However, if we take the communication, feedback and updating mechanisms seriously, we must act digital. This can be accommodated by expanding the workbench with some services, as shown in Figure 2.

It is exactly to this point that we have tied the data collection process referred to in section 2. Feedback is both direct (user comments in prose) and indirect (key click logging). Discussion is available by web linkage to an asynchronous discussion forum. All that data is instantly available to the user who can review what comments have been made at that point. Moreover, as data is collected, we can apply path detection and visualization tools to observe traffic preferences as well as use the built-in mechanisms of forum tools.

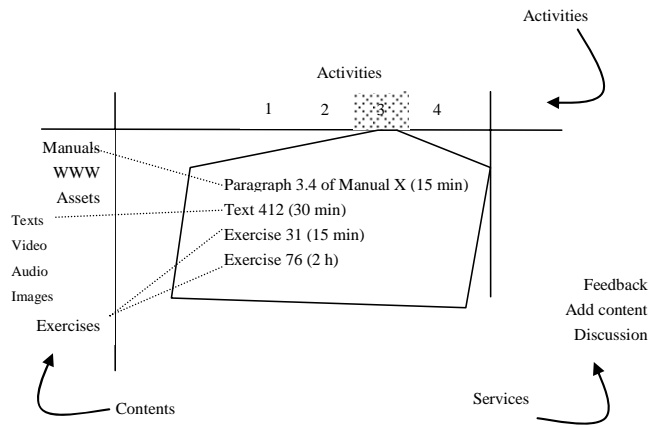

Figure 2: A tutor's workbench (with some services).

The workbench now also pertains to a material developer from the point of view of delivering the individual exercises, texts and ground-level material that will be needed in a course (photographs, videos, etc.). The add-content mechanism is useful both to the casual material developer (a tutor who takes some time to write a new presentation) and to the professional one (who will probably write the equivalent of book chapters and series of exercises).

It is also interesting to note that the new workbench is also a platform for students who can see that their activities are designed collections of (implicit) tasks that are based on a collection of assets and that numerous other ways exist for browsing or referencing the repository of the educational material. However, students also need some form of logging their progress and understanding their navigation of the course. To do so, a couple of extra services are required to allow them to observe a course map as well as see which part of the course map they have covered over a studying session (or, over a series of such sessions). This can be accommodated by still expanding the workbench with extra services, as shown in Figure 3; we also switch to referring to a "user's" workbench as the underlying platform evolves to be more generic.

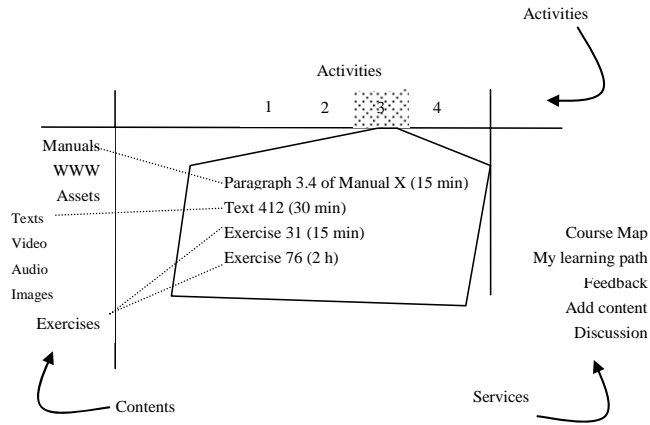

Figure 3: A user's workbench (with extra services).

More subtly, the same platform is also fit for a publisher (as a business entity) or an editor (as an academic entity). They can both play the role of the curriculum designer in delivering different (appropriately targeted) compilations of selected subsets of the educational material. That can be done using the same platform that the tutor uses to prepare an activity and that the student uses for interacting with the material and the learning community for a particular subject.

The only missing item may be the access and usage rights policy, which is mostly of an administrative nature. A simple addition in the form of "Credits" should then suffice for prototyping purposes; we now have a workable requirements specification at a high level and it is summarized in Figure 4. We also claim that this is simply a snapshot of a Learner's ODL course.

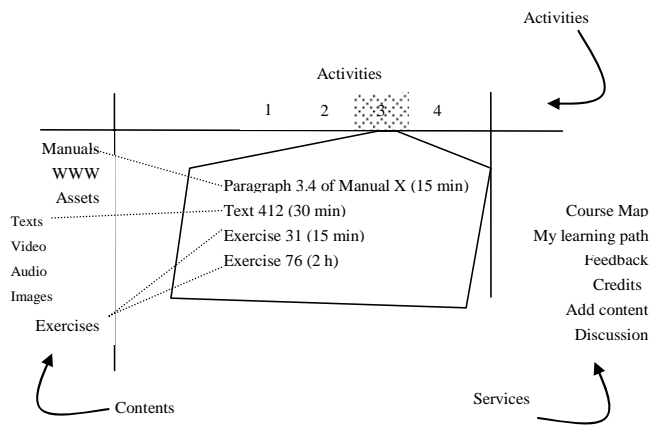

Figure 4: A user's workbench as a snapshot of a Learner's ODL course.

## 5   Designing a Learner's ODL course with a graph-theoretic model

Why do we need a mathematical model of a Learner's ODL course? We believe it was a necessary tool in our design approach because it helped formulate all important aspects of the educational process that we wanted to model and, then, seamlessly supported the semantic annotation of student activities while allowing us the convenience of knowing that graph-processing algorithms and software are a commodity.

We now elaborate on that computational model, which builds on top of some basic components as shown below.

A *learning object* is any piece of (multimedia) data or program whose purpose (intention) is to be used for learning.[3]

A *learning object* can be recursively defined as a set of learning objects.

Examples of learning objects are the following: the text of Odyssey, MS Word, Sketchpad, a video lecture, a set of multiple choice questions, a Euclidean geometry high school textbook, an MS Powerpoint presentation of organic compounds.

A *learning task* is a task whose purpose is learning.

Examples of learning tasks are the following: read, solve an exercise, write a program, practice a musical instrument, draw a picture, design a database, make a summary, think over, correct, argue for/against.

A *learning activity* is an ordered pair: *(learning object, learning task)*.

Examples of learning activities are the following:
- Write a program to add two numbers (learning task) using a C++ compiler (learning object)
- Write down (type to the computer) what you hear (the learning object is a digitized dictation) and then check the spelling errors (in fact the learning object is the set {word processor, soundtrack, speller}).

A *learning environment* is a directed labeled multigraph[4] (*LA*, *P*), where *LA* is a set (of vertices or, nodes) of learning activities and *P* is a bag (of edges) of labeled precedents.

Examples of labeled edges are the following:
- From node LA5 to node LA15 "if you found LA5 very easy to do"
- From node LA5 to node LA100 "if you found LA5 very interesting"
- From node LA5 to node LA3 "if you did not manage to complete the task of LA5 satisfactorily"

A *reference node* is (a learning activity that is) connected to all other nodes via bidirectional (unlabeled) edges.

---

[3] This can be enhanced with the usual definition and all metadata fields etc., but is not necessary here.

[4] A multigraph is a "graph whose edges are are unordered pairs of vertices, and the same pair of vertices can be connected by multiple edges". In a multigraph "edges are a bag, not a set". A bag is essentially a set with duplicates allowed. For related definitions, see http://www.nist.gov/dads/. For the rest of this paper, and for convenience reasons, we will drop the term *multigraph* and use *graph* instead.

Examples of *reference nodes* are the following:
- Dictionary (to look up a word or phrase)
- Calculator (to perform an arithmetic operation)
- On-line discussion (to communicate with a tutor or with fellow learners)

A *learning experience* (or, a *learning trip*) is a path (sequence of connected learning activities) on the learning environment graph.

A *learner's note* is a data structure attached to a specific node by a specific learner. A learner's note includes structured data fields (learner/user id, timestamp, access rights, etc.) and any (multimedia) data the learner chooses to attach (for example, files).

Examples of *learner's notes* are the following:
- The list of adjectives asked for in example B1.
- A text that criticizes the effectiveness of the learning activity (node).
- A new soundtrack of the dictation (left by a student who found the pronunciation incomprehensible).
- A comparison or a synopsis of the past 10 notes left on the current learning activity (node).

A *learning environment communication system* is a communication system (such as email, discussion forum, etc.) with content consisting only of (pointers to) learner's notes.

Examples of such content are the following:
- From a student to his teacher "Here is the list of adjectives asked for in LA5".
- From a student to all other students "I found LA12 particularly useful, you can look up my comments in the note attached".
- From a teacher to his students "Before attempting task LA112 read my note there".

A *learning activity control block* is a snapshot of the usage of all the above in the context of a particular learner. It is a data structure containing (at least) the following fields:
- learner/user id
- timestamp
- (pointer to) learning object
- (pointer to) learning task
- (pointer to) learner's note

# 6 On the potential of graph-based design

A short discussion is worthwhile here. Is a learning experience something that can be designed by a tutor? Maybe; we will delve into the issue of curriculum and course design later, and to do so we will capitalize on our graph-theoretic representation that we set out in this section. However, we will stick to the term *learning experience* to designate how a learner navigates through the learning environment. We will leave undefined, as of yet, the granularity at which a learning experience is composed and allow maximum flexibility.

In this sense, a learning experience may well be either a single-session path; for example, a learner dedicates a good solid hour to navigating the educational material along a particular line.

A learning experience may also be a sequence of such paths; for example, we usually "remember" where we stopped studying (for a short or long break), and can resume from that point. A (metaphorically speaking) concatenation of such paths delivers a longer path that can still be a learning experience.

How about path overlaps? This is a more complicated issue. It does not seem to fit well into our definition for learning experience, but, surely, we should be able to come-up with a definition that will allow us to bundle together navigation paths, associate them with the time-stamps of particular visits (see definition of *learner's note*) and use them as a representation of a coherent learning process.

The above considerations simply suggest that, after we get the initial graph-theoretic model fixed, there exist a set of computational processes that will allow us to define arbitrarily complex layers of information based on the ground data. Such information layers are, essentially, the response to the query "how does the course evolve". We now delve into one fundamental such aspect; that of the detection of learning detours.

What is a cycle in a learning environment? It is just a graph cycle and there will be *many* of them, especially if we allow our graphs to be read-write, where new edges can be added based on how an administrator (of a Learner's ODL course) establishes access rights. If we also think of the potential cycles created due to the existence of reference nodes we understand that a cycle in the learning environment will probably have limited semantic importance.

However, a cycle detected in a learning experience is a totally different issue. Such a cycle unequivocally designates a detour worth following with the intention to come back to the original point (in the worst case this suggests a top-down browsing of material).

Now, detours during learning are quite common. We occasionally need to consult additional resources while busy on a subject. Detecting cycles in learning experiences is, therefore, indicative of critical learning activities that the learner considers worth pursuing. Deleting cycles may deliver concise information about where the learner is heading (strategy) but studying them should deliver information about how the learner interacts with the material (tactics).

Having seen how even the most trivial graph-processing task can deliver useful information, we turn to see whether more complex graph-theoretic concepts and computations may be used to deliver such information.

Cliques and connected components are usually employed as a means of demonstrating graph properties that are related to localization; here, we use *localization* as a metaphor to show that some *areas* of a graph may be very close neighbours in the sense that one has to venture explicitly outside this area through very specific paths.

This is not a new concept. As a matter of fact it has been used in a very similar context, web site adaptation [6]. Automatically improving the organization and presentation of web sites based on data mining usage logs is a burgeoning scientific field and one of the approaches is based on the PageGather algorithm [7]. Therein, a clustering method, called *cluster mining*, is employed,which works on an input of user sessions, represented as sets of visited web pages (note the correspondence with learning experiences). PageGather *then* builds a graph by linking nodes (pages) with an edge whenever co-occurrence of these pages is detected across some user sessions. Page clusters (or, similar learning experiences) can then be defined using either cliques or connected components, with cliques considered to be more coherent and connected components considered to be faster to compute and easier to find.

There exist legitimate arguments about the computational cost of graph-based algorithms for inferring usage patterns [6]. However, if we can agree that our *a-posteriori* analysis of the usage (by various users) of a Learner's ODL course will be used to improve its presentation and organization in a future version -thus, we do not focus on providing immediately customizable content- [8] then these arguments are not related to our employing of the graph-based representation.

## 7 Technological support for the requirements and the design

A most basic implementation of the above, which can be also considered as view integration, can be provided by hyperlinked files of conventional office-type applications (for example, MS Excel or MS Visio), where educational assets can be grouped together in repository-type worksheets. Assets can then be drawn to compile learning activities. Such tools could possibly offer relatively smooth and short learning curves for data collection and web publishing too, as they are widely used.

For example, Figure 5 shows how MS Visio could be used to design a learning environment graph, by adding hyperlinks in nodes and edges of the complete structure. Rectangular shapes to the left correspond to assets (diagrams, text, etc.) whereas the oval shapes refer to learning tasks. When paired with assets these, then, correspond to activities, which can be linked together to create a learning environment.

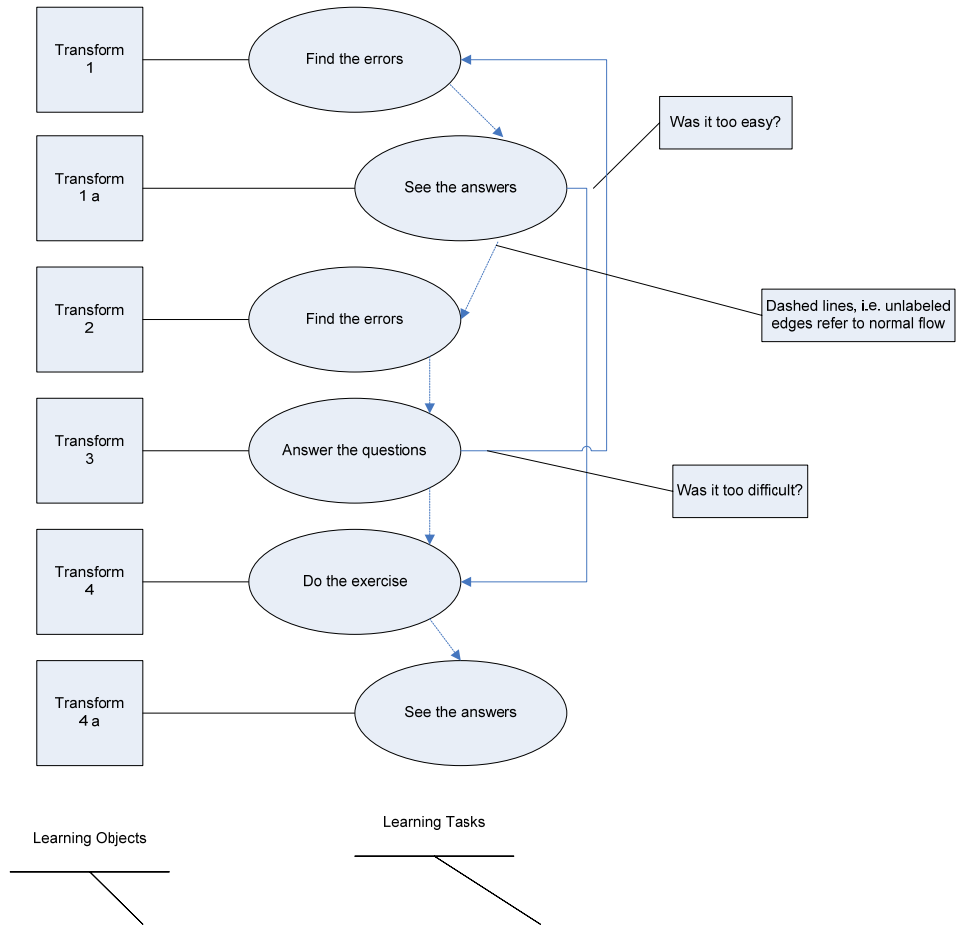

Figure 5: A snapshot of a learning environment in MS Visio.

As another example, Figure 6 shows how MS Excel could be used to design a learning environment. The leftmost column contains an indicator of the asset type. At the same line, the learning activity is composed by an asset and by a learning task (allowing, of course, for some terseness in representation: when no task is shown for a text asset, the implicit task is "read"). Indentation can be used to designate priorities and preferences; this allows transitions between activities to be tagged by attributes such as "was it interesting?", just as shown in Figure 5.

| Learning Task | Learning Object | | | | |
|---|---|---|---|---|---|
| The divide-and-conquer approach ... | | | | | |
| read | 2.3.1 The divide-and-conquer approach | | | | |
| | | ... the first two paragraphs | | | |
| write | Think how you would apply the above principle to …. | | | | |
| read | 2.3.1 The divide-and-conquer approach | | | | |
| | | ... the next paragraph | | | |
| write | You might want to rethink your previous answer | | | | |
| | Think about the following details | | | | |
| exerc | | How do you split in two a sequence that has an odd number of elements? | | | |
| exerc | | How do you decide that a sub-problem is "small enough"? | | | |
| exerc | | | Is there an oprimal number of sequences? | | |
| read | | 2.1 Insertion sort | | | |
| read | | 2.2 Analyzing algorithms | | | |
| observe | | Presentation by MIT OCW Algorithms Lecture 01 | | | |
| read | 2.3.1 The divide-and-conquer approach | | | | |
| | | ... the next three paragraphs | | | |
| programming | Write a program for mergesort (do not test it) | | | | |
| exerc | What kind of input do you think you need for testing? | | | | |
| www | | See an applet that demonstrates the mergesort algorithm | | | |
| www | | See a collection of sorting algorithms | | | |
| exerc | | | Can you argue which of the above algorithms are divide-n-conquer? | | |

Figure 6: A snapshot of a learning environment in MS Excel.

After one settles on the issue of the implementation of the basic model, the issues of linkage with external resources (such as discussion rooms, and other related communication-oriented applications) and operational deployment must be addressed. At that point, one can opt to start integrating different technology offerings (having, of course, to address the overhead of inter-application communication) or adopting a generic platform approach that will allow for customization to retro-fit the implementation of the model as well [4, 9]. The latter approach can be more scalable (for example, portal offerings by commercial organizations) but the analysis to decide on such an investment may be too difficult to carry out effectively (hidden costs can surface quite easily and the steepness of the learning curve for developers may be expensive to estimate) [10, 11, 12].

However, there also exist some in-between approaches; in these approaches one may decide to use building blocks based on generic digital object identification schemes, such as DOI (http://www.doi.org) and expect that third-party providers (for example, one's university) will supply the naming space, and couple these identification schemes with generic object ensemble builders, such as Fedora [13] and SCORM [14], which accommodate a disciplined format of digital object creation and manipulation.

Web usage mining for learning cycle detection is more complicated, however. To-date the most prominent web mining applications have been demonstrated in e-commerce, which are fundamentally different compared to e-learning applications [15]. Still, as far as the underlying technologies are concerned, there is much room for technology transfer.

The analogy is clear: if there is a procedure for a course developer to track and identify paths as sequences of visits to specific nodes (we liberally use the paradigm described in Section 5), then this can lead to semantic information on the best possible adaptation of a course for a particular learner, or even to the reconstruction of the course (enriching some paths or removing –apparently- useless content). Such flexibility is primarily important in delivering network-based distance learning education but can also be a useful tool in any e-learning setting.

The technical challenge is how to relate the relatively flat structure of web log files with the apparently deep structure of learning experiences (therein, we note again the introduction of cycles in experience paths). Our technical approach is to specify the course multi-graph in advance[5], also confirming published experience [16] about the difficulties of developing a data pre-processing environment in a distance learning educational domain.

A coarse example of these concepts is shown below.[6] Figure 7 demonstrates the course multi-graph structure, as specified by the course designer (actually, it is a view of the multi-graph where, for the sake of conciseness, we have only included learning activities). Figure 8 shows a learner's path during a single learning session in the course, with nodes being numbered according to the relative order of visit.

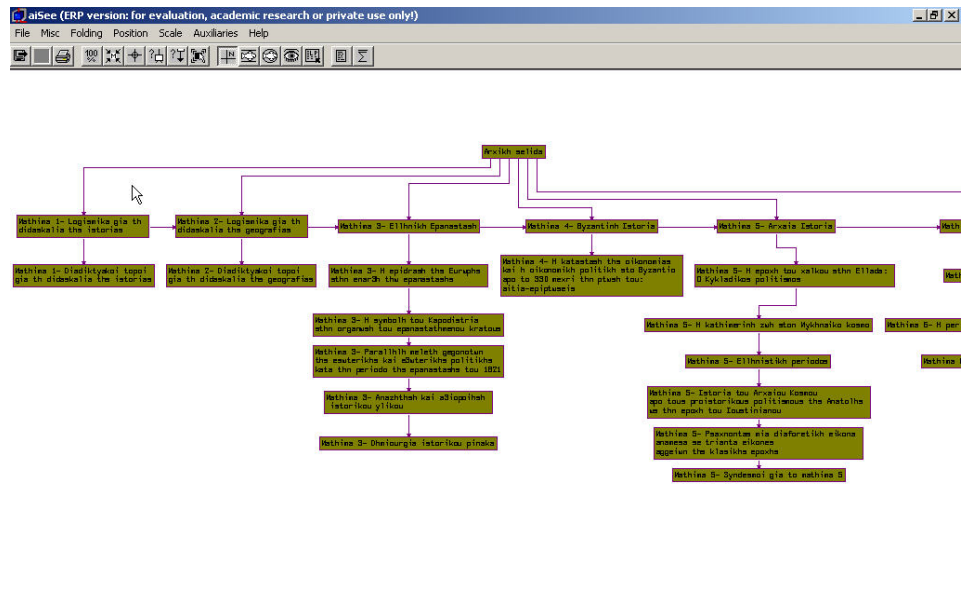

Figure 7: A view of the course multi-graph.

---

[5] We embed *php* scripts interfacing to a *mySql* database in the course's *html* code.

[6] Graph visualization employs the *aiSee* tool (http://www.aisee.com). All text is Greek in Latin format.

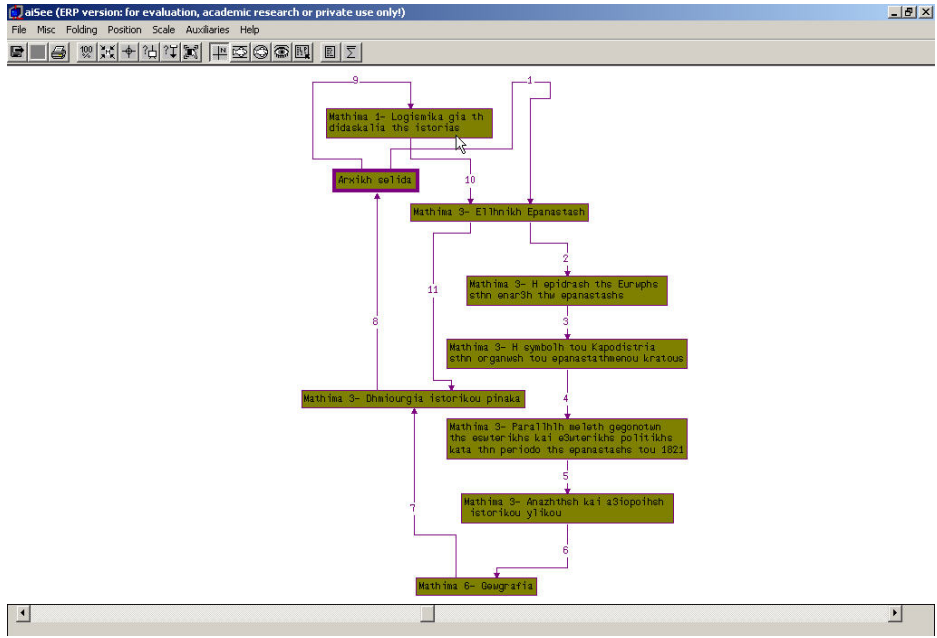

Figure 8: A visitor's path through the course.

We currently use a slight variant of the above mechanism to implement the note-passing mechanism between students and tutors. However, for this particular aspect, we are also investigating the usefulness of open-source asynchronous discussion forum systems (and the extent to which they might accommodate the multi-graph specification as opposed to programming it from scratch).

Our organization has lately completed a major transition to a common commercial portal platform (IBM WebSphere and Lotus Domino) and has initiated the installation and operation of an open-source digital asset management system (DSpace) as well as a professional SCORM-compliant authoring tool (Trivantis Lectora). Deploying the newly-developed courses on those platforms will facilitate the production and sophisticated analysis of log files, as outlined above. We stress that it is imperative, for any operational phase at a university scale, to be able to as much utilize existing technology and applications. This may sometimes mean that we must strive for our models to be designed with their implementation in mind (which is typical engineering but not engineering as being taught). To that end we are also experimenting with the possibility of developing path detection as a web-service to be provided by a third party at the course deployment level as opposed to on-line log file analysis.

## 8   Implementation and deployment vis-à-vis the application domain

We have not yet addressed the above implementation issues holistically. A primary reason is that the whole concept is complex to communicate to all stakeholders at once, since it represents a shift from the current status. The least difficult factor may well turn out to be the student view: indeed editors and tutors need to change their approach to designing. A second reason is that substantial resources are required for "indoctrination" at several levels: at the educational process level, at the tool level, at the support level, and beyond.

We have thus chosen the usual approach of pilots: by closely working with a diverse range of tutors we hope to fine-tune the model representation as regards its communication to third parties. We expect that when non-conventional users of educational technology (unfortunately but not surprisingly, that also includes IT instructors) manage to feel at ease with some development services, then these services (tools and support) may be ready for prime time.

To achieve this consensus we are also addressing one problem at a time. As opposed to a bundling approach, we develop, in-parallel, several (template) Learner's ODL courses. Each one aims to focus on implementing the basic model along a particular axis. For instance, one axis is concerned with using a Learner's ODL course as an asset repository across collaborating groups (this is akin to indexing already developed learning objects, learning activities and reference nodes of the model presented in Section 5). Another axis is concerned with providing the asset repository and guiding the development of a Learner's ODL course towards the development of new learning activities with clearly identified learner actions – see Figure 6 for an example (this is akin to developing learning objects, learning activities and reference nodes of the model presented in Section 5, with special emphasis on the role of learning tasks in the context of a learning activity). A third approach aims at enhancing already deployed web-based material with a collaborative learning mechanism – this is done by designing and implementing navigation path detection and visualization (in essence, implementing the aspects of a learning experience, a learner's note and a learning environment communication system of the model presented in Section 5, as shown in Figure 7 and in Figure 8).

Addressing one problem at a time allows us to test different aspects of the concept with different groups. Admittedly, using the same test groups for various implementations, or the same implementation aspects with various group, would probably result in a more robust methodology but we have chosen our approach in order to receive as much feedback as soon as possible. In this sense, rudimentary improvements are always incorporated and each course builds on next. The really important stake [5] is to convince as many potential authors (tutors) as possible to participate and to devote time to the collective material and course development process.

Streamlining a process that is known to work at the laboratory level so that it can work efficiently at mass production level is well known in established areas of engineering (productionizing). However, in educational technology we are often faced with the situation where small experiments (always successful by their very nature: a motivated educationalist is bound to succeed where directly involved) fail to scale up to any significant degree. Moreover, large-scale "educational" technology is more often than not poor in substance.

Setting up a software product line is a challenging undertaking itself [17], yet in the realm of educational content we usually need to address the issues of the platforms and the content as a combination, making the whole effort even more difficult. While addressing our key goal, designing a process targeted at the average tutor, we have also the following engineering issues to address, so that all stakeholders are well served:
- Facilitate editing and archiving.
- Allow for multiple dissemination channels.
- Take into account the scalability of results depending on hardware and experience available (else: utilize better tools when available but do not depend on them for a result).
- Allow for unforeseen medium-term changes in technology.

## 9  Conclusions

Open Universities, such as the Hellenic Open University, are conceived with a large number of students as a target and must therefore use mature, available and reliable technologies for their educational materials. Sometimes this can become very restrictive, leading to books and TV alone, but gradually the technology tide starts making alternative offerings more cost-effective. We have gradually embarked on e-learning initiatives, spanning from virtual classrooms, to discussion forums and, now, on to the mass-scale development of complementary on-line material.

The piecewise development and deployment of Learner's ODL courses, as set out above, has had a significant result as of now: it has reinforced that practical relevance of the underlying theoretical model, since we have been able with relatively low effort to publish those model aspects that were deemed important for each user group. Scaling the number of groups, the type of groups and the size of groups are all actions that have to be scheduled as we are moving out of the laboratory and into the users' offices.

In this paper we have argued about the importance of a simple underlying model that will capture the software and knowledge engineering aspects of the educational process. Why did we *not* use a different model? Actually we did. The Excel example was our first implementation attempt at attracting fellow tutors to the didactical merits of explicitly stating learning tasks and expected time for related activities. Note that these very tutors may well be excellent when addressing an audience; it is their skills at developing distance learning material that we aim to further develop. So, after all, we first used a tabular model, which was the easiest to communicate. But the approach matured in a way not unlike database design. While most of us will probably still find it easier to discuss database implementation issues in terms of tables, especially when discussing such things with practice-honed non-IT people, we still have to revert to conceptual modelling to describe the semantic richness of our applications.

Thus, taking into account that we need to also address the needs of tutors with limited IT skills, the careful selection of tools for the initial compilation and development of learning activities is always a key factor in our decisions.

In that sense, we believe that our key contribution is the bridging of design richness and implementation practicalities in the context of a very large scale project of distance learning digital educational material. Of course, we do need to find a way to bridge the gap with large-scale top-down efforts, such as Learning Design[7]. However, we feel that similar situations will be common in the context of almost all organizations developing digital educational content and our experience might shape the eventual shift in how other learning institutions conduct their educational tasks at a distance.

---

[7] http://www.imsglobal.org/